# Digital ASIC Design with Ongoing LLMs: Strategies and Prospects

Maoyang Xiang, Emil Goh, T. Hui Teo*
Engineering Product Development
Singapore University of Technology and Design
*Corresponding author: tthui@sutd.edu.sg
These authors contribute to the work equally.

The escalating complexity of modern digital systems has imposed significant challenges on integrated circuit (IC) design, necessitating tools that can simplify the IC design flow. The advent of Large Language Models (LLMs) has been seen as a promising development, with the potential to automate the generation of Hardware Description Language (HDL) code, thereby streamlining digital IC design. However, the practical application of LLMs in this area faces substantial hurdles. Notably, current LLMs often generate HDL code with small but critical syntax errors and struggle to accurately convey the high-level semantics of circuit designs. These issues significantly undermine the utility of LLMs for IC design, leading to misinterpretations and inefficiencies.

In response to these challenges, this paper presents targeted strategies to harness the capabilities of LLMs for digital ASIC design. We outline approaches that improve the reliability and accuracy of HDL code generation by LLMs. As a practical demonstration of these strategies, we detail the development of a simple three-phase Pulse Width Modulation (PWM) generator. This project, part of the "Efabless AI-Generated Open-Source Chip Design Challenge," successfully passed the Design Rule Check (DRC) and was fabricated, showcasing the potential of LLMs to enhance digital ASIC design. This work underscores the feasibility and benefits of integrating LLMs into the IC design process, offering a novel approach to overcoming the complexities of modern digital systems.

# Introduction

The landscape of modern digital integrated circuit (IC) design is marked by an ever-increasing complexity, a trend vividly illustrated by the evolution of processors in mobile phones. These processors, which are fundamental to the device's operation and encompass crucial components such as the baseband and image signal processor, have undergone rapid evolution. They now deliver enhanced speed, efficiency, and functionality, all within ever more compact dimensions. This complexity not only demands extensive resources for design and verification but also significantly extends the development cycle, emphasizing the critical need for tools and methodologies that can simplify the IC design flow. Addressing the complexities posed by modern digital systems, therefore, becomes imperative for sustaining innovation and meeting the burgeoning demands of technological advancements.

Recent years have seen remarkable advancements in the field of artificial intelligence, particularly in the development of Large Language Models (LLMs). These models, trained on vast datasets to understand and generate human-like text, have demonstrated potential across various fields including the code generation. In the domain of digital IC design, LLMs offer the promise of automating the generation of Hardware Description Language (HDL) code, potentially streamlining the design process. The ability of HDL code generation could significantly reduce the manual effort involved in writing and debugging HDL, offering a novel approach to tackling the complexities of IC design.

However, the application of LLMs in IC design is not without challenges. Key issues include the generation of syntax errors and difficulties in accurately interpreting high-level circuit semantics. These problems can lead to inaccuracies in the generated HDL code, undermining the reliability of the design process and potentially introducing significant delays in development timelines.

In response to these challenges, this paper aims to introduce and discuss strategies for leveraging modern language models in digital ASIC design. Our focus is on overcoming the obstacles presented by syntax errors and semantic interpretation, with the goal of improving the accuracy and reliability of HDL code generation through LLMs. We illustrate the application of these strategies through the development of a simple three-phase Pulse Width Modulation (PWM) generator, which not only showcases the practical application of our approach but also highlights our contribution to the "Efabless Unveils Winners of 3rd AI-Generated Open-Source Chip Design Challenge." The success of this project underscores the feasibility of using LLMs in IC design and points to a future where they play a significant role in streamlining the design process.

The organization of this paper is as follows: Section 2 presents a literature review of LLMs and their application in generating HDL code, providing context and background for our strategies. The following section details our proposed strategies for leveraging LLMs in digital ASIC design, addressing the identified challenges. Section 4 discusses the simulation of the generated code, demonstrating the efficacy of our approach. Finally, this research delves into a broader discussion of the potential and future implications of LLMs in digital IC design, offering insights into how these models can transform the design landscape.

# Literature Review

Traditional Hardware Description Language (HDL) coding practices are acknowledged for their labor-intensive nature and susceptibility to errors, presenting substantial challenges within the design process. This section amalgamates current strategies and methodologies aimed at augmenting the efficiency of the digital ASIC design flow.

The utilization of high-level languages for the generation of HDL code represents an area of ongoing exploration. Among the contributors to this domain, Amjad et al. (2019) delineated a methodology that enables the translation of code from the Signal language to Verilog, exemplifying a novel technique to expedite the HDL code generation process [1]. Concurrently, Tatsuoka et al. have integrated High-Level Synthesis (HLS) with physically aware logic synthesis technology, facilitating early identification and resolution of congestion issues within the design phase [2]. Furthermore, Shinya Takamaeda-Yamazaki and colleagues have introduced Pyverilog, an open-source toolkit for RTL design analysis and Verilog code generation from Python, thereby simplifying the development process and allowing designers to focus more on the chip functionality itself [3]. Despite the advantages, these approaches possess limitations; synthesizers or translators often act as opaque entities, and the resultant HDL code may exhibit behaviors, particularly concerning timing, that deviate from the designer's expectations.

In parallel, the rapid advancement of artificial intelligence, exemplified by OpenAI's GPT series, has showcased the potential of Large Language Models (LLMs) in understanding and generating human-like text. This has extended into the realm of HDL code automation. Dehaerne et al. (2023) introduced a pioneering framework for Verilog autocompletion using deep learning, thereby streamlining the code writing and verification processes [4]. Ahmad et al. (2023) explored the use of LLMs for automatically rectifying security-relevant flaws in hardware designs crafted in Verilog [5]. Moreover, research by Thakur et al. (2023) underscored the proficiency of LLMs in automating hardware design through the generation of Verilog code, emphasizing the efficacy of finely-tuned models in producing syntactically accurate code [6]. Thakur et al. further investigated the capacity of LLMs to generate HDL models by integrating insights from compiler tools, enhancing code precision [7]. Meanwhile, Emil et al. purpose the digital IC implementation by fine-tuning of Mistral-7B LLM model for plain language to HDL/RTL [8].

Despite the optimistic trajectory for employing LLMs in ASIC design, numerous challenges persist. A critical issue is the necessity for LLMs to precisely comprehend and generate domain-specific technical content. Moreover, the reliability and veracity of LLM-generated outputs are crucial, as inaccuracies could culminate in expensive design setbacks.

The prospect of applying LLMs in ASIC design is indeed promising, yet it is accompanied by significant challenges. Primarily, these models must accurately understand and produce specialized technical content. Ensuring the reliability and integrity of the outputs generated by LLMs is vital, given that inaccuracies may result in costly design errors. In subsequent chapters, we will delve into methodologies for leveraging large predictive models to generate more precise and reliable integrated circuit designs.

# Method

This section highlights crucial considerations for enhancing the quality and effectiveness of HDL code generated by LLMs in the domain of digital circuit design. It underscores the importance of role specification, hierarchical design methodologies, and robust error feedback mechanisms.

## Role Specification

During its training phase, the LLM extensively assimilates information across diverse levels and styles. This process of assimilation introduces significant uncertainties within the code generation phase, particularly evident in the production of low-quality and stylistically inconsistent HDL code. Consequently, it is imperative to define a precise role and establish a consistent coding style for the LLM before engaging in the generation of corresponding HDL code. For example, the LLM could be designated as a specialist in integrated circuit (IC) design, or its coding style could be explicitly articulated.

To illustrate this point, consider the use of ChatGPT-4 to generate a [3-to-8 decoder](). The provision of a specific role, such as an IC design specialist, enhances the quality of the generated code. Notably, in the generated example, there is an enable signal within the module, which aligns closely with low-power IC design methodologies. This specification not only facilitates the generation of efficient Verilog code but also adheres to a robust Verilog coding style.

However, it is important to note that without such role specification, the initial generated module may present issues such as glitches and multi-drive conflicts. Although modern synthesizers can address these issues, they are still considered undesirable practices in Verilog coding. Therefore, specifying the role of the LLM in advance significantly improves the coherence and technical integrity of the generated HDL code.

## Hierarchical Digital System Description

Contemporary digital circuit systems are complex, large-scale entities. The intricate nature of system design inherently encompasses substantial diversity, which introduces significant challenges when employing LLMs to generate designs. Specifically, LLMs must autonomously determine the myriad details of the digital systems they generate, while IC designers may lose control over the granular details of these systems.

In this context, a more effective utilization of LLMs in circuit design involves hierarchical digital circuit design. This approach mandates that IC designers explicitly define or review the overarching framework, while the LLMs operationalize each segment of the circuit within this pre-approved structure. Such a strategy enhances the enforcement of functional constraints and the verification of functional completeness.

To illustrate, consider two examples where ChatGPT-4 was used to assist in the design of a Stopwatch, serving to compare the efficacy of this method in digital circuit generation. The system in question is a simple digital circuit, specifically a four-digit stopwatch that provides a stable display via a four-digit digital tube. Both design approaches adhered to the same constraints, specifically the clock frequency input for each module.

The code generated from these designs was synthesized using Vivado 2023.1. The results, as is shown in Table 1, indicated that the hierarchical design approach not only utilized fewer logic resources but also achieved shorter synthesis times, underscoring the benefits of this method in digital circuit design.

Table 1 : Synthesis Report of Generated Verilog Code with Different Approachs

| Verilog Code LLM Approach | Slice LUTs | Slice Registers | Synthesis Time (s) |
| --- | --- | --- | --- |
| With hierarchical | 70 | 54 | 14 |
| Without hierarchical | 95 | 78 | 16 |

## Verilog Code Error Feedback

HDL code generated by large language models is likely to have minor but fatal syntax errors. Therefore, it is very necessary to establish a feedback mechanism for generating HDL. The establishment of this feedback mechanism mainly includes feedback of syntax error information and behavior simulation error information. The following is a list of several common errors in LLM generated Verilog code.

### Mismatch begin-end block;

In the example, an always @(*) begin block is mistakenly closed with a } instead of an end. This leads to a syntax error because the code block is not properly terminated according to Verilog syntax rules.

```
always @(*)begin
}
```

### Mismatch case-endcase block;

Similar to the mismatch in begin-end blocks, a case block must be closed with endcase. The example shows a case block that starts correctly but ends with end instead of endcase.

```
case()
   `STATU1:;
   default:;
end
```

### Multi-Drive Issue;

A multi-drive issue occurs when multiple drivers (sources) attempt to assign a value to the same wire or reg at the same time, leading to conflicts and unpredictable behavior. In the provided example, out[1:0] is driven by different values within the same always @(*) block based on the case condition. This can result in conflict, especially in synthesis, where the last assignment in a procedural block may override previous ones, potentially causing glitches or logic errors in the output.

```
always@(*)begin
   out[1:0] = 2'b00;
   case (in)
```

```
    `STATU1: out[1:0] = 2'b00;
    `STATU2: out[1:0] = 2'b01;
    //Other statement;
  endcase
end
```

Ambiguous Clock;

Ambiguity in the clocking signal arises when the sensitivity list of the always block contains more than one edge-triggered event, which can lead to unpredictable behavior. In the example, the out will be driven on the rising edge of clock and falling edge of rstn, which result in a conflict;

```
always @(posedge clock, negedge rstn)begin
  if (!rstn) begin
  //...
  end else begin
  //...
  end
  out <= 1'b0;
end
```

Overall, by integrating precise role specification, hierarchical design principles, and effective error feedback mechanisms, the use of LLMs in digital circuit design can be optimized to produce high-quality, reliable, and efficient HDL code, thereby addressing both the challenges and leveraging the capabilities of artificial intelligence in complex engineering applications.

# Results

## Design the three phase PWM module

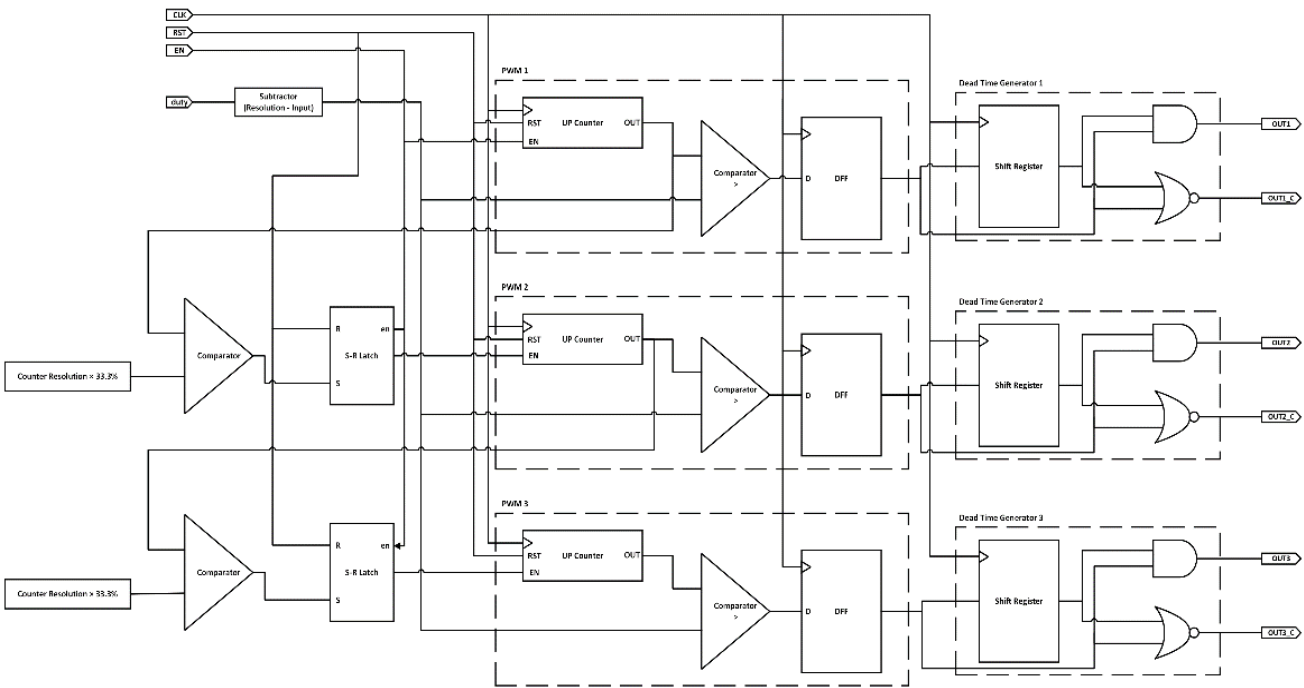

Figure 1 : Three phase PWM generator block diagram

To validate the strategies to generate Verilog code, they were employed to develop a sample design, a three-phase PWM generator, which was designed to demonstrate. It was also submitted for Efabless' AI-Generated Open-Source Design Challenge to extend the Caravel SoC peripherals. The three-phase PWM circuit is designed to generate 3 PWM waveforms with its complementary signal (6 PWM waveforms altogether), each 120 degrees out of phase with the others. By modifying the duty cycle input signal, the output waveforms' duty cycle can also be adjusted. With dead time, the PWM waveforms will not be toggled HIGH and LOW simultaneously, which could cause a short circuit. This ensures safe operation in applications when driving power electronics.

The three-phase PWM design can be broken down into the following components, as is shown in Figure 1:
- Control
  - Signal
    - Clock: Main clock signal to drive circuit;
    - Reset: To reset the circuit to the default state;
    - Enable: To activate and deactivate the circuit;
  - Duty Cycle Control
    - Duty Cycle Input: Represents the desired PWM duty cycle;
    - Subtracts the input duty cycle with the counter resolution, inverting the input to the PWM comparators;
  - PWMs, each consisting of the following:
    - Up Counter: 8-bit counter (256 resolution), automatically resets to 0 after hitting 255;
    - Comparator: Compares the counter value with the subtracted duty cycle value to produce the PWM waveform;
    - D Flip Flop (DFF): To filter out unwanted transitions;
  - Phase Control (2 sets), each set consists of the following:
    - "One-third" Comparator: Compares the previous PWM's counter output with 33.3\% of the counter resolution. As such, the comparator will only produce a HIGH signal when the previous counter reaches 85/256;
    - S-R Latch: Used to sustain the HIGH signal produced by the "one-third" comparator;
  - Dead Time Generators, each consisting of the following:
    - Shift Register (4 DFF cascaded together): delays each PWM output by 4 clock cycles;
    - AND gate: Produces the main PWM output;
    - Produces the complementary PWM output;

## Simulation

The simulation is done using iVerilog and waveform viewed on GTKWave. The duty cycle is increased from 0% to 25%, 50%, and 75%. As expected, the design generates 3 PWM with 3 complementary signals. The overall waveform can be viewed in Figure 2.

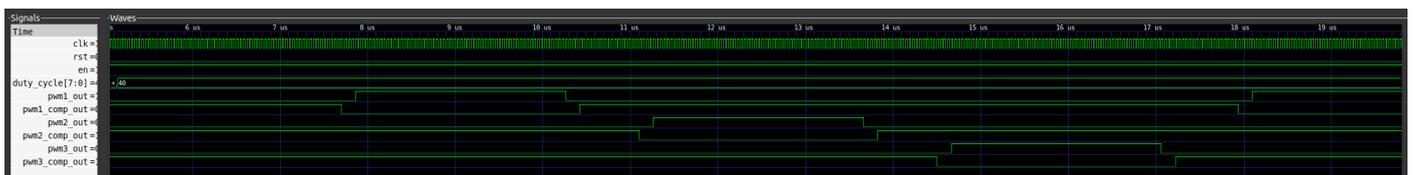

Figure 2 : Simulation result for PWM generator

## Discussion

LLMs have begun to be explored for their capability in generating Verilog code. This application area, however, presents several unique challenges and opportunities which will be discussed in this section.

### Difficulty with Timing Diagrams:

Timing diagrams are crucial in hardware design as they visually represent the relationship between different signals in a circuit over time. These diagrams are essential for understanding the sequence and coordination required in circuits, such as Serial Peripheral Interface (SPI) communications. LLMs currently struggle to interpret these diagrams directly and translate them into Verilog code. This difficulty arises because LLMs primarily handle textual and syntactic processing and lack the ability to directly analyze visual data. To effectively generate Verilog code based on timing diagrams, additional layers of interpretation or pre-processing might be necessary to convert visual information into a text-based format that LLMs can understand.

### Quality of Training Data:

The efficacy of an LLM in generating accurate and syntactically correct Verilog code heavily depends on the quality of the training data. Currently, there is a shortage of high-quality, stylistically consistent Verilog code in the datasets used to train these models. This results in LLMs often reproducing the same errors or poor coding practices found in their training data. Furthermore, discrepancies in how different synthesizers handle Verilog code can exacerbate these issues. Synthesizers might accept syntactic expressions that do not align with the Verilog-2003 standard or other established norms, leading to inconsistencies and potential errors in generated code. Addressing these challenges would require curating and possibly sanitizing training datasets to ensure that LLMs learn from clean, standard-compliant examples.

### Potential in Code Generation:

Despite these challenges, LLMs demonstrate significant potential in automating the generation of Verilog code. The progress in this area is notable—from generating nonsensical code snippets to producing coherent and functionally viable digital systems. This evolution indicates that with the right training data and further refinement of model architectures and training techniques, LLMs could become powerful tools in digital circuit design. They could help automate tedious aspects of code documentation, suggest design modifications, and potentially innovate in circuit design through generative techniques.

In conclusion, while LLMs face challenges in generating Verilog code, particularly with interpreting complex visual data and dealing with inconsistent training datasets, their potential to enhance and streamline the process of digital circuit design is undeniable. Future advancements could involve better integration of visual data processing capabilities and improvements in data quality for training, which would significantly bolster the practical utility of LLMs in electronic design automation (EDA).

## Acknowledgments

We would like to thank SUTD-ZJU IDEA Visiting Professor Grant (SUTD-ZJU (VP) 202103, and SUTD-ZJU Thematic Research Grant (SUTD-ZJU (TR) 202204), for supporting this work.

# Appendix

- 3-to-8 Decoder with ChatGPT4: https://chat.openai.com/share/1514435d-3d7b-4be9-8653-55cff94db6ee
- Stopwatch with ChatGPT4 (hierarchical): https://chat.openai.com/share/0574c38c-71fd-442f-b941-f5613c208a02
- Stopwatch with ChatGPT4 (non-hierarchical): https://chat.openai.com/share/23cebdd1-31f3-48e8-98eb-187ba3176bbb
- Efabless Competition Link: https://efabless.com/news/efabless-unveils-winners-of-3rd-ai-generated-open-source-chip-design-challenge
- Efabless Verilog PWM Design Tutorial: https://chat.openai.com/share/ff55a56d-56fa-400e-878d-bc47a241caec
- Efabless Three-Phase PWM Design Implementation: https://chat.openai.com/share/6900a303-3c33-4b63-9cd5-74ca69348593